\documentclass[prl,twocolumn,showpacs,floatfix,aps]{revtex4}

\usepackage{graphicx}

\begin{document}
\input epsf


\title{Search for E$_{2g}$ phonon modes in MgB$_{2}$ single
crystals by point-contact spectroscopy}

\author{Yu. G. Naidyuk\thanks{e-mail: naidyuk@ilt.kharkov.ua},
I. K. Yanson, O. E. Kvitnitskaya}

\affiliation{B. Verkin Institute for Low Temperature Physics and
Engineering, National Academy  of Sciences of Ukraine,  47 Lenin
Ave., 61103,  Kharkiv, Ukraine}

\author{S. Lee, and S. Tajima}
\affiliation{Superconductivity Research Laboratory, ISTEC, 1-10-13
Shinonome, Koto-ku, Tokyo 135-0062, Japan}

\date{\today}

\begin{abstract}
The electron-phonon interaction in magnesium diboride MgB$_{2}$
single crystals is investigated by point-contact (PC)
spectroscopy. For the first time the electron coupling with
E$_{2g}$ phonon modes is resolved in the PC spectra. The
correlation between intensity of the extremely broad E$_{2g}$
modes in the PC spectra and value of the superconducting gap is
established. Our observations favor current theoretical models for
electron-phonon mediated superconductivity in MgB$_{2}$ and they
better match the harmonic phonons model.

\pacs{74.25.Fy, 74.80.Fp, 73.40.Jn}
\end{abstract}

\maketitle


{\it Introduction}. During the two year rush in study of
superconductivity in MgB$_2$ the electron-phonon coupling has been
debated as a main source of the Cooper pairing. Many of
theoretical calculations report similar findings: the E$_{2g}$
phonon modes, corresponding to the in-plane distortions of the
boron hexagons, have an enormous coupling to electrons and are
responsible for high T$_c$
\cite{AP,Yildirim,Kortus,Liu,Bohnen,Kong,Golubov,Choi}.
Correspondingly, the main feature of the calculated
electron-phonon interaction (EPI) function $\alpha^2F(\omega)$ is
a dominant maximum at the energy of  stretching E$_{2g}$ modes
within 60-70\,meV \cite{Liu,Bohnen,Kong,Golubov,Choi}. However, in
spite of the great progress in computation of $\alpha^2F(\omega)$,
which is calculated even for various parts of the Fermi surface
\cite{Golubov,Choi}, direct experimental verification for the
$\alpha^2F(\omega)$ behavior is still lacking.

The importance of the E$_{2g}$ modes is evidenced by the Ra\-man
spect\-roscopy studies, which demonstrate strongly damped and
broad optical modes with the E$_{2g}$ symmetry
\cite{Bohnen,Renker,Quilty,Raman}. Additionally, an anomalous
large broadening of the E$_{2g}$ phonon modes along the $\Gamma$-A
direction is recently observed in single crystals by inelastic
X-ray scattering \cite{Shukla}. It is shown that the dominant
contribution to the linewidth is due to a strong electron-phonon
interaction (EPI). But both the foregoing experiments deal with
phonons and not directly with the $\alpha^2F(\omega)$ function.

The two well known fundamental experimental methods for
determining $\alpha^2F(\omega)$ are point-contact (PC) and
tunneling spectroscopy. Both of them were successfully utilized
for observation and study of superconducting gap(s) (see, e. g.,
Refs.\,\cite{Szabo,Bugos,Naidyuk,Gonnelli,Schmidt,Giubileo,Iavarone}).
Nevertheless, only a few papers report the search for the phonon
structure and EPI in PC \cite{Bobrov,Yansmgb} or in tunneling
spectra \cite{Dyachenko}.

Bobrov {\it et al.} \cite{Bobrov} communicated reproducible
maxima, including a maximum around 60\,mV, in the PC spectra for
different contacts prepared on the c-axis oriented MgB$_2$ films.
In the forthcoming paper Yanson {\it et al.} \cite{Yansmgb}
stressed that in the superconducting state the phonon structure
observed in the PC spectra was due to the energy dependence of the
superconducting order parameter \cite{YansonSC,Beloborod}. For
strong coupled superconductors this, so called, elastic term
\cite{Beloborod} exceeds the inelastic EPI contribution to the PC
spectra determined directly by $\alpha^2F(\omega)$ \cite{Kulik}.
Consequently, to separate or to distinguish the elastic and
inelastic effects measurements of PC spectra of MgB$_2$ in the
normal state are required. However, the resolution in the
point-contact spectroscopy (PCS) is limited by temperature and at
$T\geq T_c\simeq 40$\,K it is above 20\,mV. On the other hand, the
MgB$_2$ thin films have a large critical field above 25\,T
\cite{Buzea} to drive them in the normal state at helium
temperature.

This paper reports the investigation of MgB$_2$ single crystals by
PCS. The advantages of the single crystal as compared to the films
in the context of the foregoing discussion are a few times lower
critical field, permitting superconductivity to be almost suppress
by moderate field along the c-direction \cite{SLee}, the best
sample quality (lower residual resistance or larger mean free
path) and the possibility of anisotropy study.

The goal of the investigations is to recover the EPI function for
MgB$_2$ by means of PCS in order to elucidate the mentioned above
issue about role of E$_{2g}$ modes in superconductivity of this
compound.

{\it Experimental details}. The single crystals of MgB$_2$ were
grown in a quasi-ternary Mg-MgB$_2$-BN system at a pressure of 4 -
6 GPa and temperature 1400-1700\,$^o$C for 5 to 60\,min. In
optimal conditions shiny yellow-colored single crystals with the
large side of 0.2 - 0.7\,mm were yielded. The details of the
sample preparation are given elsewhere \cite{SLee}. The estimated
resistivity at 40\,K is about 1\,$\mu\Omega$\,cm and the residual
resistivity ratio RRR=5$\pm$ 0.1. A sharp superconducting
transition in resistivity is around 38.1-38.3\,K with $\delta T_
c$(10-90)\% = 0.2 - 0.3\,K.

The PC`s were established {\it in situ} at low temperatures by
touching of a Cu or Au electrode by tiny piece (scrap) of the
MgB$_2$ single crystal. That is because of the small size and the
irregular shape the MgB$_2$ crystal was used as a "needle". We
always try to align the c axis of the MgB$_2$ single crystal with
an applied magnetic field and the contact axis was along the
largest side, presumably, perpendicular to the c axis in most
cases. Due to lack of special micro-mechanics the accuracy of
orientation was held by eye. No more control of the sample
orientation was possible during the measurements. Therefore,
information about the contact axis orientation with respect to the
main crystallographic direction was gained from the analysis of
the gap features.


{\it Results and discussion}. Study of nonlinear conductivity of
metallic PC`s  allows the EPI function $\alpha^2F(\omega)$ to be
recovered directly by measuring the second derivative of the $I-V$
characteristic (see, e. g., Ref. \cite{Kulik,Yanson}):
\begin{equation}
\label{pcs} R_0^{\rm -1}\frac{{\rm d}R}{{\rm d}V}= A
R_0^{-1/2}\alpha_{\rm PC}^2(\epsilon)\,F(\epsilon)|_{\epsilon={\rm
e}V} ,
\end{equation}
where $A$ is the constant, $R={\rm d}V/{\rm d}I$, $R_0$ is the
contact resistance at zero bias, $\alpha^2_{\rm PC}F(\omega)$ is
the PC EPI function, which differs from the Eliashberg EPI
function by the presence of factor $K=1/2(1-\theta/\tan\theta)$,
where $\theta$ is the angle between initial and final momenta of
scattered electrons [for the transport and Eliashberg EPI
functions the corresponding factors are: $K=(1-\cos\theta)$ and
$K$=1, respectively].
\begin{figure}[t]
\begin{center}
\includegraphics[width=7.5cm,angle=0]{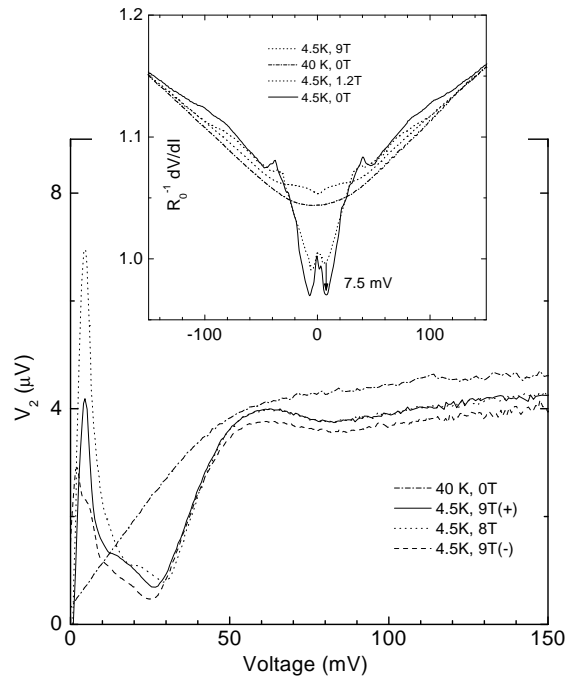}
\end{center}\vspace{-20mm}
\caption[] {PC spectra for the MgB$_2$-Cu contact
($R_0=7.2\,\Omega$) at $T=4.2$K in a magnetic field 8\,T (dotted
curve) and 9\,T (solid curve). The dashed curve is for the
opposite polarity at 9\,T. Above 30\,mV the curves practically
coincide. The dash-dotted curve is recorded at zero field at
40\,K. The modulation signal $V_1(0)$ is 3\,mV. Inset: the
d$V$/d$I(V)$ curves for the same contact. The bottom two curves
are for zero and 1.2\,T field and 4.2K.} \label{fig1}\vspace{-6mm}
\end{figure}
From Eq.(\ref{pcs}) $\alpha_{\rm PC}^2(\epsilon)\,F(\epsilon)$ can
be expressed via the measured rms signal of the first $V_1 \propto
$d$V/$d$I$ and the second $V_2\propto $d$^2V/$d$I^2$ harmonics of
a small alternating voltage superimposed on the ramped $dc$
voltage $V$:
\begin{equation}
\label{pcs1} \alpha_{PC}^2(\epsilon)\,F(\epsilon)=
\frac{2\sqrt{2}}{A}R_0^{1/2}\frac{V_2}{V_1^2} .
\end{equation}
It should be emphasized that to discard the influence of the
superconducting peculiarities we have analyzed the PC spectra
measured at a maximal magnetic field (as high as 9\,T) which is
close to the upper critical field in the MgB$_2$ single crystal
\cite{SLee}) and/or above T$_c\leq$40\,K. The common features for
all the spectra measured at low temperatures are the residual
superconducting structure (the sharp maximum near zero bias) and
the continuously raising structureless background above 25-30\,mV.
For a number of curves a smoothed maximum (rather a hump) between
40-80\,mV was resolved. The examples of the measured dependences
with the maximum are shown in Figs.\,1-3. First of all we should
stress that the structure in the spectra shown above 30\,mV is
robust with varying the magnetic field in the range 6-9\,T, while
the superconducting structure below 30\,mV depresses drastically
by a magnetic field (see Fig.\,1). Secondly, above 30\,mV the
d$^2V$/d$I^2(V)$ curves are reproducible for both polarities of
the applied voltage. This is in support of the phonon nature of
the maxima(um) in d$^2V$/d$I^2(V)$. From the minima position in
the d$V$/d$I$ curves in the insets of Figs.\,1 and 2 measured at
zero field and low temperature we can estimate a gap value which
is around 7.5\,meV and 2.7\,meV, respectively. Considering the
well fixed by PCS and tunneling fact \cite{Gonnelli,Iavarone} that
a gap of about 7\,mV corresponds to the 2-D $\sigma$ bands and a
large gap is seen for the direction of the boron planes, we
suggest that the spectra in Fig.\,1 are measured presumably along
the defined direction of the boron sheets. Additional support that
we have really observed a large gap in this case is provided by
the magnetic field data. A field about 1\,T only slightly modified
the gap structure in d$V$/d$I$ (Fig.\,1, inset), while according
to Ref.\cite{Gonnelli}, this field is sufficient to suppress  the
small gap. In this connection maxima at about 60\,mV in the
spectra in Fig.\,1 are naturally to connect with the broadened
E$_{2g}$ phonon modes, which predominate in $\alpha^2F(\omega)$
for the $\sigma$ band \cite{Golubov}.
\begin{figure}[t]
\begin{center}
\includegraphics[width=8cm,angle=0]{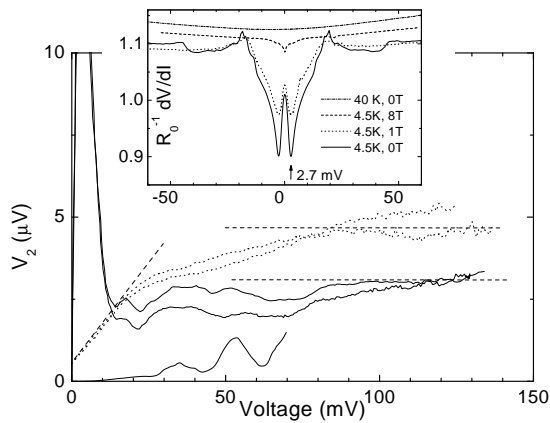}
\end{center}\vspace{-5mm}
\caption[] {PC spectra for the MgB$_2$-Cu contact
($R_0=1.5\,\Omega$, $V_1(0)$=3.2\,mV) at $T=4.2$\,K and $B$=8\,T
(solid curves). The dashed curves are recorded at zero field and
40\,K. Couple curves are shown for the both polarities of the
applied voltage. Dashed straight lines show position of a kink and
saturation in the PC spectra. The bottom curve is the low energy
part of the isotropic $\alpha^2F(\omega)$ \cite{Golubov} smoothed
by the experimental resolution about 6\,mV. Inset: the
d$V$/d$I(V)$ curves for the same contact.}
\label{fig2}\vspace{-0mm}
\end{figure}

Unlike the spectrum in Fig.\,1, the spectrum in Fig.\,2 with a
small gap about 2.7\,meV is attributed to the 3-D $\pi$ bands and
is governed by the directions out of the boron planes.
Correspondingly, the shallow maxima close to 30 and 50\,mV reflect
bulk (isotropic) phonons, and their correlation with the first two
maxima in the phonon DOS \cite{Renker} or $\alpha^2F(\omega)$
\cite{Bohnen,Kong,Golubov} is evident. We should also emphasize
that the intensity of the spectrum in Fig.\,1 is about two times
higher than that in Fig.\,2, which, according to Eq.\,(2), results
in about 5 times larger value for $\alpha^2F(\omega)$ due to the
difference in the PC resistance $R_0$. This is also in line with
an expected larger strength of EPI in the boron planes.
\begin{figure}[t]
\begin{center}
\includegraphics[width=8cm,angle=0]{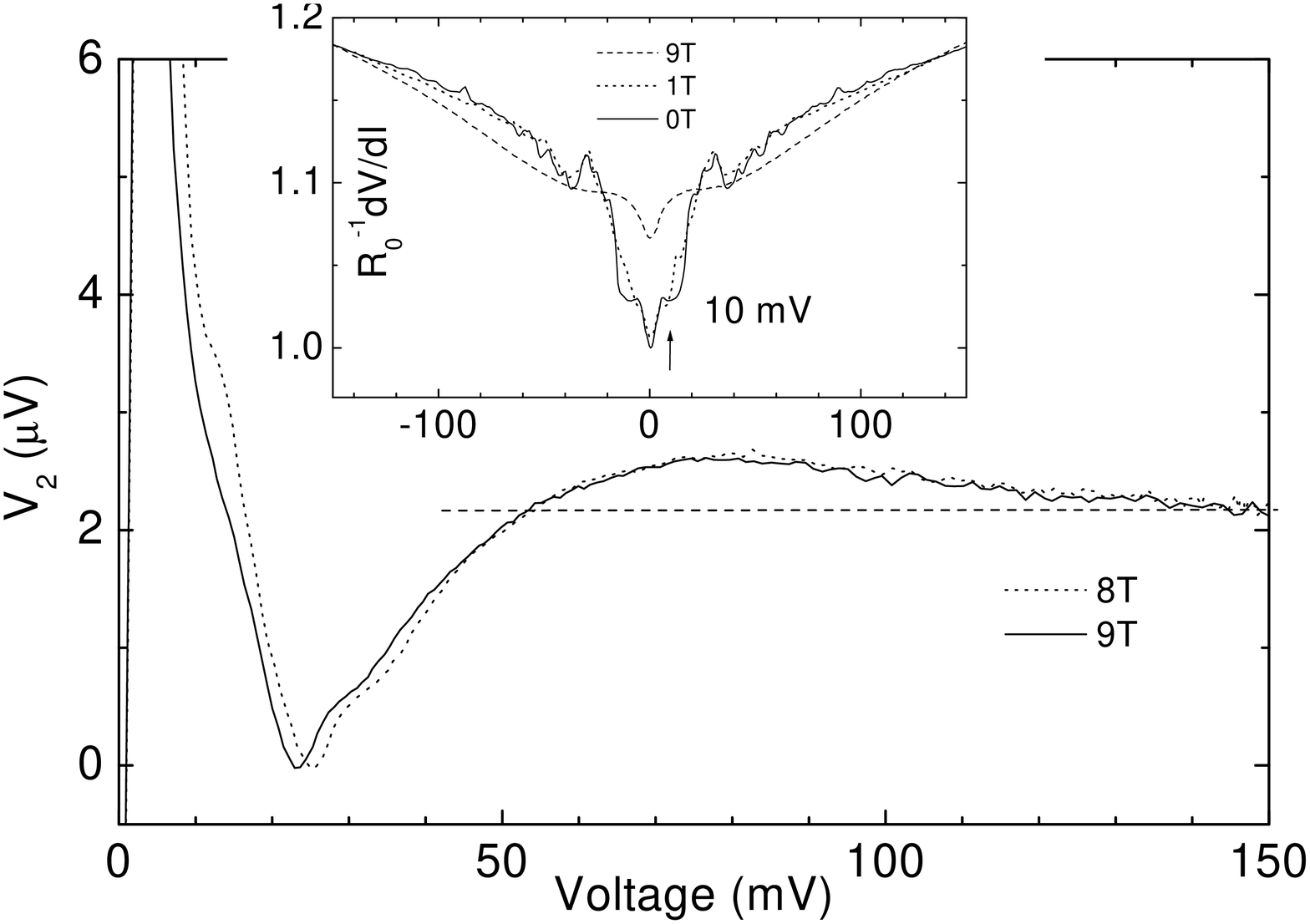}
\end{center}\vspace{-5mm}
\caption[]{PC spectra for MgB$_2$-Cu contact ($R_0=7.5\,\Omega$,
$V_1(0)$=2.5\,mV) at $T=4.2$K in a magnetic field 8\,T (dashed curve)
and 9\,T (solid curve). Curves are almost the same for the both
polarities of the applied voltage. Inset:
d$V$/d$I(V)$ curves for the same contact. } \label{fig3}
\vspace{-3mm}
\end{figure}

Take a note of the spectra in Figs.1 and 2 measured above $T_c$
around 40\,K. Both of them exhibit a buckle (kink) at the position
of the maximum in the corresponding low temperature spectra as
well as the saturation above 80\,mV. This is an additional
confirmation of the phonon caused structure in the curves shown.

Fig.\,3 depicts an extra example of the PC spectrum with a smooth
maximum around 75\,mV which also correlates with the dominating
maximum in  many theoretical $\alpha^2 F(\omega)$ as well as with
the Raman data (see Fig.\,4(b)). Note that though d$V$/d$I$ is
deviated from the BTK shape because of a minimum at $V$=0 and
over-gap maxima, the so-called "gap" minima are observed here at
about $|V|\simeq$10\,mV. Therefore, we attribute these spectra to
the $\sigma$ band.

Further comparison of our data with the Raman spectrum shows that
the Raman maximum (Fig.\,4(b)) is narrower and shifted to a higher
energy with respect to that in the PC spectrum in Fig.\,4(a). Note
that the Raman spectroscopy studies are restricted to the
Brillouin zone center, and the Raman maxima, in general, should
not correspond, e. g., to the maxima in the phonon DOS or
especially in $\alpha_{\rm PC}^2 F(\omega)$, which reflects
phonons with a large (maximal) momentum.
\begin{figure}[t]
\begin{center}
\includegraphics[width=7cm,keepaspectratio=true]{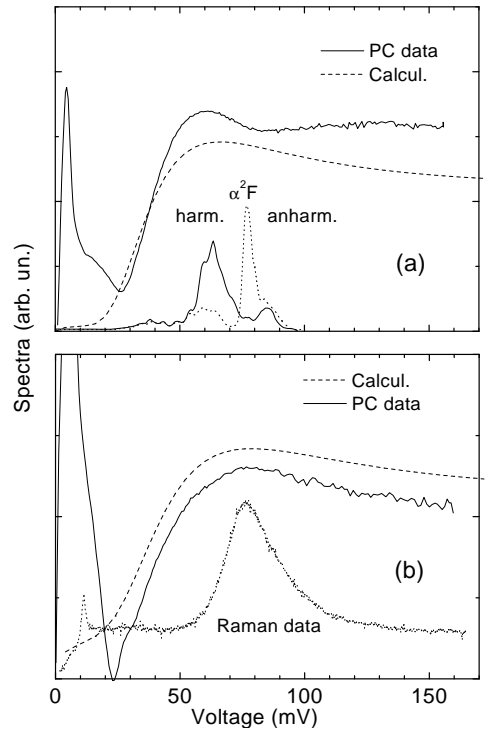}
\end{center}\vspace{-5mm}
\caption[]{(a) PC spectrum from Fig.1 transformed into
$R_0^{-1}$d$R/$d$V\propto V_2/V_1^2$ in comparison with the
calculated one according to Eq.(3) for the Lorentz-shaped peak at
60\,mV and a width of 2\,mV. The bottom curves show $\alpha^2
F(\omega)$\cite{Choi} for harmonic and anharmonic phonons. (b) The
PC spectrum from Fig.\,3 in comparison with the calculated one
according to Eq.\,(3) for the Lorentz-shaped peak at 70\,mV and a
width of 15\,mV. The bottom curve is depolarized Raman spectra of
MgB$_2$\cite{Quilty} at 21\,K.} \label{fig4}\vspace{-5mm}
\end{figure}

According to the inelastic X-ray scattering data \cite{Shukla} an
anomalous broadening of the E$_{2g}$ modes is found to occur along
the $\Gamma$-A direction with a width $\delta E$ =10 - 17\,meV and
phonon dispersion between 60 and 70\,meV. This energy range hits
the maximum in the spectra in Fig.\,1, however, it is fairly
broad. Using the width of the E$_{2g}$ modes along the $\Gamma$-A
direction we estimate the phonon lifetime as
$\tau\propto\hbar/\delta E \simeq (4-7)\cdot 10^{-14}$s. Taking
into account the upper value of the sound velocity $s\leq 10^4
$m/s \cite{Fil} this leads to the phonon mean free path of the
order of the lattice constant. Therefore, the energized electron
generated nonequilibrium phonons accumulate in the contact region,
giving rise to (i) a large background signal \cite{Kulik} and (ii)
a thermalization of phonons directly in the PC and a transition to
a thermal regime\cite{Kulik}. In the latter case the phonon
features in the PC spectra are considerably smeared out.

The PC EPI spectrum in the thermal regime supposing that the
phonon part of the resistivity is small compared with the residual
resistivity is given by Kulik \cite{Kulik}:
\begin{equation}
R_0^{-1}\frac{{\rm d}R}{{\rm d}V}(V)= C\int_0^{\infty}\frac{{\rm
d}\omega}{\omega}\,\alpha^2F(\omega)\,S(eV/\hbar\omega) ,
\label{EPIth}
\end{equation}
where $C$ is the constant and $S(x)$ represents a smeared around
$x$=1 step with a shallow maximum at $x$=1.09. We have calculated
the PC spectrum by (3) for $\alpha^2F(\omega)$ where the E$_{2g}$
mode was simulated by the Lorentz-shaped peak. Comparison with the
experimental curves is shown in Fig.\,4. The results describe
fairly well the spectrum of the contact from Fig.\,3 and shows
that the spectrum in Fig.\,1 has a more distinct maximum, as
expected in the thermal regime. The position of this maximum
corresponds to the main peak in $\alpha^2F(\omega)$ calculated for
the harmonic phonon model \cite{Choi}. This correlates with the
statement by Shukla {\it et al.} \cite{Shukla} according to which
the anharmonic contribution to the phonon linewidth is much
smaller compare to the electron-phonon one.

Turn to the EPI inelastic contribution in the PC spectrum, which
is about 10\% of the total PC resistance (see Fig.\,1, the inset).
it is a few times higher than for the MgB$_2$ thin
films\cite{Bobrov}. However, it is of the same order as that for
the non-superconducting transition metal diborides \cite{NaidMeB},
where the parameter EPI is rather low $\lambda\lesssim 0.1$.
Conventional explanation considers a deviation from the ballistic
transport, leading to a decrease in the intensity of the PC
spectra \cite{Kulik,KulYan}. However, due to the strong variation
of EPI in MgB$_2$ on the Fermi surface and the fact that the large
EPI is confined to a small volume in the ${\bf k}$-space along the
$\Gamma$-A direction \cite{Choi} the PC EPI function $\alpha_{\rm
PC}^2 F(\omega)$ can be vastly different compared with the
thermodynamic $\alpha^2 F(\omega)$ one. Therefore, to elucidate
the details of EPI in MgB$_2$, the calculation of $\alpha_{\rm
PC}^2 F(\omega)$ with the mentioned $K$-factor accounting for
strong anisotropy of EPI on the Fermi surface and the phonon
lifetime effects is very desirable.

{\it Conclusion}. We have measured anisotropic PC spectra in the
single crystal of MgB$_{2}$. By virtue of the strong EPI and the
short phonon lifetime a very likely close-to-thermal-regime state
develops for the most PC`s, which hinders receiving detailed
information about the spectral EPI function. However, certain of
the PC`s display a spectrum, which corresponds to EPI in the
$\sigma$ band, and exhibits damped broad maximum just above
60\,meV caused by the E$_{2g}$ phonon modes. The $\pi$ band
spectrum reveals only lower lying phonon modes. Accordingly, for
the first time the experimental manifestation of the anisotropic
interaction of electrons with E$_{2g}$ phonons is presented. Our
observation is also in consistent with the harmonic model of EPI
in MgB$_{2}$.

{\it Acknowledgments}. The work was supported by the National
Academy of Sciences of Ukraine and by the New Energy and
Industrial Technology Development Organization (NEDO) in Japan. We
are indebted to A. Golubov and O. Dolgov for providing the
$\alpha^2 F(\omega)$ data. The investigations were carried out in
part with the use of the equipment donated by the Alexander von
Humboldt Foundation (Germany).


\end{document}